\documentclass[%
 aip,
 amsmath,amssymb,
 reprint,%
]{revtex4-1}

\usepackage{graphicx}
\usepackage{dcolumn}
\usepackage{bm}

\usepackage[utf8]{inputenc}
\usepackage[T1]{fontenc}
\usepackage{mathptmx}
\usepackage{etoolbox}

\makeatletter
\def\@email#1#2{%
 \endgroup
 \patchcmd{\titleblock@produce}
  {\frontmatter@RRAPformat}
  {\frontmatter@RRAPformat{\produce@RRAP{*#1\href{mailto:#2}{#2}}}\frontmatter@RRAPformat}
  {}{}
}%
\makeatother

\begin{document}

\title[SQUID-based interferometric accelerometer]{SQUID-based interferometric accelerometer}

\author{I. Khomchenko}
\affiliation{Digital Engineering Center, Skolkovo Institute of Science and Technology, 30 Bolshoi Boulevard, bld. 1, Moscow 121205, Russia}

\author{P. Navez}
\affiliation{Department of Physics, Loughborough University, Loughborough, LE11 3TU, United Kingdom}%

\author{H. Ouerdane}
\affiliation{Digital Engineering Center, Skolkovo Institute of Science and Technology, 30 Bolshoi Boulevard, bld. 1, Moscow 121205, Russia}%

\date{\today}

\begin{abstract}
Optics and more recently coherent matter waves enabled inertial sensors such as accelerometers and gyroscopes to reach high levels of resolution and sensitivity. As these technologies rest on physical phenomena that require particular setups and working conditions such as, e.g., kilometers of optical fibers or ultralow temperatures, their application range is limited because of lack of portability. Here, we propose a path forward considering a superconducting quantum interference device (SQUID) to detect and measure acceleration, using electronic interferometry. The basic idea is \emph{not} to use a SQUID as a magnetometer in acceleration measurement setups, but as an accelerometer. The operation of such an accelerometer rests on the ability of the Cooper pairs to record their wave function phase change as the device is subjected either to a transverse acceleration or vibrations. We provide numerical evidence for the feasibility of SQUID-based accelerometers that can be used for transverse acceleration and oscillatory motion measurement.
\end{abstract}

\maketitle

Interferometry-based sensors have numerous practical applications such as, e.g., geosensing, inertial navigation, local gravitational field measurements, gravitational wave detection, and quantum gravity \cite{Igel2007GJI,Geiger2011,Leveque2021,Will2006LRIR,Wu2019,Margalit2021}. Initiated first in the field of optics, considerable improvements have been achieved by employing matter waves of ultra-cold Bose-condensed atoms: as the device sensitivity increases linearly with the total particle mass energy \cite{Bouyer}, passing from photons ($\sim$ 1 eV) to atoms ($\gtrsim$ 10 GeV) results in an increase of sensitivity by ten orders of magnitude \cite{Carnal1991PRL,Keith1991PRL,Kasevich1992APB,Schaff2014RDNC,Barrett2014CRP}. In order to measure the Earth rotation using the Sagnac effect, kilometers of optical fiber are needed in contrast to a millimeter ring trap for atoms. While both setups present advantages, the former has the inconvenient to be cumbersome while the latter uses laser cooling technology in a vacuum chamber, which also bears some important practical constraints \cite{sagnac1913ether,sagnac1914effet,post1967sagnac}. 

To extend the range of situations for which interferometry-based technologies may be used, it is natural to consider electrons as ``phase change recorders'' in a quantum circuit made of superconductors. The main limit of this approach is that the sensitivity increases by only six orders of magnitude with respect to photon-based devices as the electron mass is 0.511 MeV, but this can be compensated by the benefit of using more portable microelectronics technology, which does not require the system to be cooled down to ultra-cold temperatures. In a superconductor, a Cooper pair would record the phase along its trajectory in wires at cryogenic temperatures. The simplest device for the purpose is a superconducting quantum interference device, or SQUID \cite{BookChap22004,squid2006}. Note that using London's prediction that a rotating superconductor produces a magnetic field throughout its interior, the intensity of which is proportional to the angular velocity \cite{London1950,Tate1990,Mach,Fischer2001}, makes already a SQUID an interesting device for the measurement of the rotation which differs from the Sagnac effect. 

Here, we theoretically demonstrate the possibility for superconducting devices to be used as convenient and accurate inertial sensors. While SQUIDs have been used before for gravimetry and acceleration measurement \cite{Chan1987,Book2001,Dittus2001,Rzhevskiy2020}, they were operating as magnetometers; here, the idea is to employ SQUIDs as accelerometers rather than as magnetometers. With this approach, a SQUID can be designed so that its loop area can be very small, which mitigates issues such as external noise effects \cite{BookChap72004} and allows reaching high sensitivity to acceleration ($\sim 10^{-10}~{\rm m}\cdot{\rm s}^{-2}/\sqrt{\rm Hz}$). In this work, we study the phase and current variations of SQUIDs experiencing either a constant transverse acceleration, or oscillatory forces due to vibrations. Our basic system being a circuit made of superconductors, we adapt a formalism used for ultra-cold atoms \cite{Storey1994,navez2016matter} to the case when the Cooper pairs' motion is restricted by the confining potential of the superconducting wires. For a transverse acceleration $a$, we recover the simple approximation of phase change $\delta \phi =m v at^2/\hbar $ for a Cooper pair of mass $m$ of constant linear velocity $v$ during time $t$. \cite{Bouyer} The geometry of the SQUID plays an important role in the final account of the measurable phase difference. For this reason, we determine the electrical response to a constant acceleration for both the square and ring geometries. 

Our paper is organized as follows. Using a semi-classical approach, we start with the basic principles of our model considering dc SQUIDs with ring and rectangular geometries for constant transverse acceleration. We establish a relationship between acceleration, supercurrent, and a geometrical form factor. Next, we provide an analysis of the sensitivity to acceleration and show that a SQUID with a small loop surface, if used as an accelerometer, can provide highly accurate measurement data. Finally, we extend our analysis in the frame of the time-dependent Gross-Pitaevskii formalism to evaluate the magnitude of possible deviations of a Cooper pair along its trajectory as the SQUID is subjected to a transverse acceleration.

\begin{figure}
\includegraphics[width=.5\textwidth]{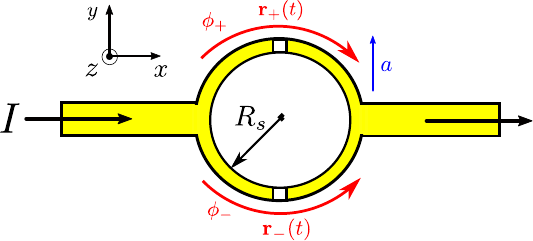}
\includegraphics[width=.5\textwidth]{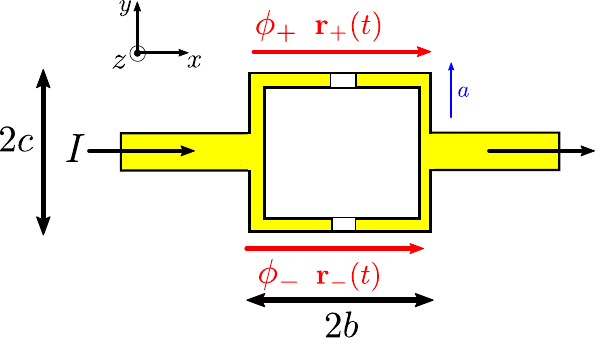}
\caption[SQUID pictures]{Representation of a SQUID with two geometries: (a) ring and (b) rectangular. The Cooper pair current and the related phase shift $\delta\phi$ are caused by the transverse acceleration ${\bf a}$. The $\pm$ symbols indicate the trajectories $\mathbf{r}_{+}$ and $\mathbf{r}_{-}$ and the phases $\phi_{\pm}$ along the upper and lower arms respectively.}
\label{Squids}
\end{figure}

Let us consider typical SQUID setups such as those schematized in Fig.~\ref{Squids}. The net current of Cooper pairs (of mass $m$ and charge $q=2e$ with $e$ being the elementary charge), induced by the dc Josephson effect and beyond which a voltage is noticeable, reads:
\begin{equation}
I=2I_{\rm c} \cos(\delta \phi/2)
\label{IdcJ}
\end{equation}
with $I_{\rm c}$ being the effective critical current, $\delta \phi = 2\pi \Phi_B/\Phi_0$ the phase shift generated inside the loop by an external magnetic field ${\bf B}$, and $\Phi_0=h/2e$ is the superconducting flux quantum. In absence of external magnetic field, subjecting the SQUID to a transverse acceleration causes the phase of the Cooper pairs condensate to change. This is the effect we propose to exploit for acceleration measurement. So, assuming that the phase shift is caused by an acceleration $a$ along the $y$-axis perpendicular to the direction of current $I$ ($x$-axis), the resulting phase is obtained from the difference of the phase accumulation in each arm $\pm$ (see Fig.~\ref{Squids}) expressed as:\cite{Storey1994,navez2016matter}
\begin{eqnarray}
\delta \phi &=& \phi_+-\phi_- \quad \quad \mbox{with} \quad
\phi_\pm =\int_0^{t_{\pm}} \frac{m {\bf \dot r}_{\pm}^2(t)}{2\hbar} {\rm d}t
\label{eq:phi}
\end{eqnarray}
where ${\bf \dot r}_{\pm}(t)$ is the superconducting fraction's center of mass velocity in each arm of the SQUID, and $t_{\pm}$ the traveling times along the trajectories ${\bf r}_{\pm}(t)$ from the beginning to the end of each arm. Quite generally, we may assume that the total current through the wire is constant. Under these conditions, provided that the cross section of the wires is constant, the absolute value for the velocity is constant as well. This observation allows to determine unambiguously the phase difference if the wire geometry is known. 

Considering first the ring geometry characterized by the radius $R_{\rm s}$, the Cooper pairs' trajectories are described by the vectors ${\bf r}_{\pm}$ at time $t$: ${\bf r}_{\pm}(t)= (R_{\rm s}\cos(\Omega t),\pm R_{\rm s}\sin(\Omega t) + at^2/2 ,0)$, with $\Omega=v/R_{\rm s}$ being the effective angular velocity, and $v$ the drift velocity; so the total phase accumulated along the loop may thus be expressed as: 
\begin{equation}\label{dphir}
\delta \phi 
=-2\frac{m}{\hbar}
\int_0^{\pi/\Omega}\!\!\!\! R_{\rm s} a \Omega t \cos(\Omega t){\rm d}t
=f\frac{a}{I}
\end{equation}
\noindent where $f=4m R_{\rm s}^2I/(\hbar v)$ 
is a form factor associated to the SQUID geometry. For the rectangular geometry, $f=m 2c(2b+c)I/\hbar v$.\cite{SuppMat} The substitution of the speed $v$ by the electrical current in each arm $I/2=(2e)nvd\lambda$ through the device yields:
\begin{eqnarray}
\label{geomfac}
f= 
\begin{cases}
8m R_{\rm s}^2(2e)nd\lambda/\hbar & {\rm Ring} \\
4mc(2b+c)(2e)nd\lambda/\hbar & {\rm Rectangle}
\end{cases}
\end{eqnarray}
with $n$ being the Cooper pairs density, $d$ the cross-section diameter of the wire, and $\lambda$ the London penetration depth. 

For gravity measurements, the cause of acceleration is the gravitational field, in which case ${\bf a}={\bf g}$. Using the gravitational potential energy in the action, we determine the phase difference from the expression: 
\begin{eqnarray}
\delta \phi &=& \phi_+-\phi_- \quad \quad \mbox{with} \quad
\phi_\pm =\int_0^{t_{\pm}} \frac{m {\bf g}\cdot{\bf  r}_{\pm}(t)}{\hbar} {\rm d}t
\label{eq:phi2}
\end{eqnarray}
Assuming the gravitational field along the $y$-axis, a Cooper pair does not experience the same gravitational field along its trajectory in the two arms. With ${\bf r}_{\pm}(t)= (R_{\rm s}\cos(\Omega t),\pm R_{\rm s}\sin(\Omega t),0)$ for the ring geometry, the phase difference reads $\delta \phi=fg/I$, which is necessarily identical to Eq.~\eqref{dphir} as a consequence of Einstein's equivalence principle.

\begin{figure}[h]
  \includegraphics[width=.45\textwidth]{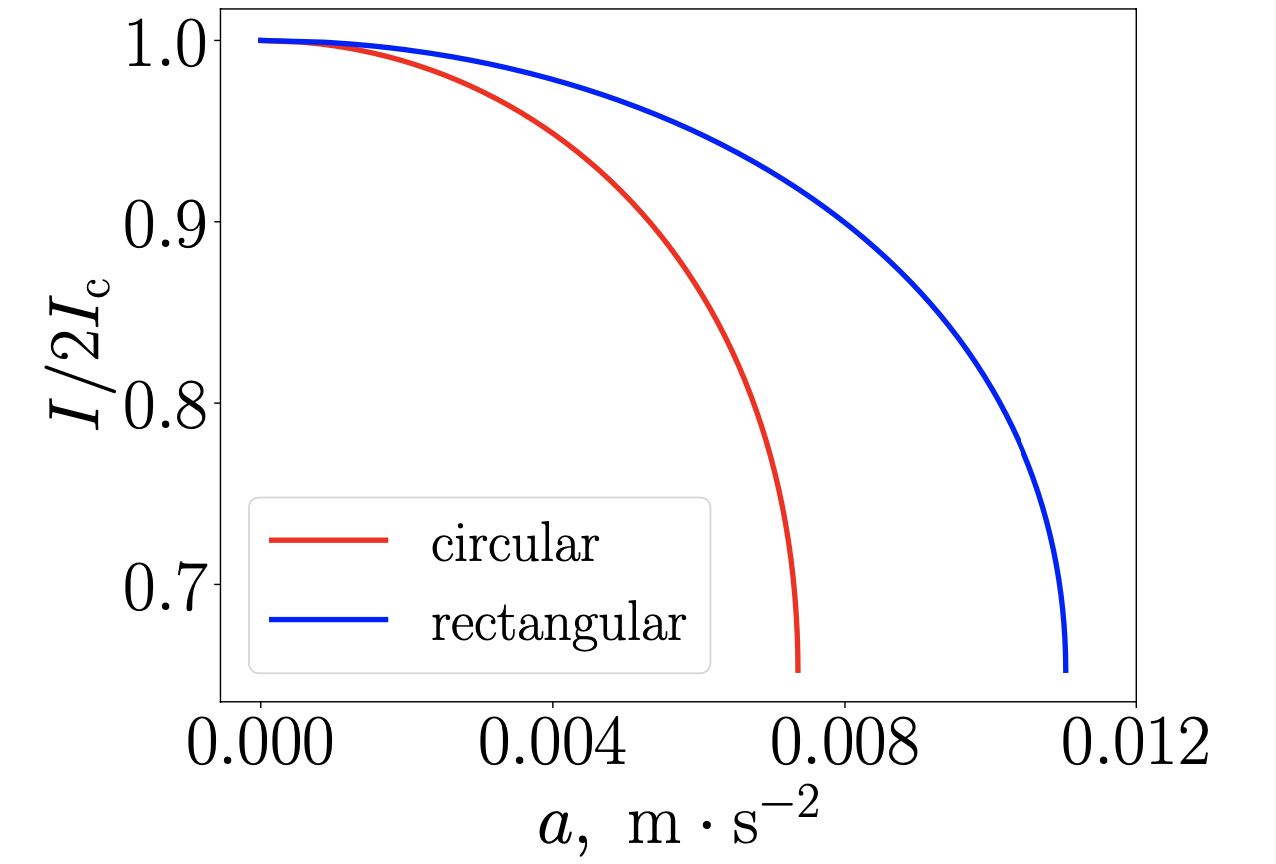}
  \caption{Relative current variation $I/2I_{\rm c}$ in each arm as function of acceleration $a$. The red line is for the rectangular geometry with $b = c = 300~\mu$m, and the blue line for the circular geometry with $R_{\rm s}=300~\mu$m. For both geometries, the cross section diameter is $d = 10~\mu$m, the Cooper pairs concentration $n = 1.0 \times 10^{23}$ cm$^{-3}$, the critical current $I_{\rm c} = 0.5 \mu$A, the penetration depth $\lambda = 5\times 10^{-8}$ m.}
  \label{ssc}
  \vspace*{-0.4cm}
\end{figure}

From Eq.~\eqref{IdcJ}, we obtain a simple expression of the Cooper pair current:
\begin{equation}
a = 4\sqrt{I_{\rm c}(2I_{\rm c}-I)}/f
\label{eq:ic}
\end{equation}

\noindent for weak phase shifts due to a small acceleration $a$, i.e. when $a < I/f$. Note that another weak current regime exists when the phase is locked to $\delta \phi=fa/I=\pi$ but does not display any observable voltage as the current in the arms are not close to $I_{\rm c}$. We may now determine the transverse acceleration $a$ to which the SQUID-based accelerometer is subjected for both geometries by measurement of the total current $I$ in the circuit. Figure~\ref{ssc} shows the ratio $I/2I_{\rm c}$ as a function acceleration $a$ for the two considered geometries and a given Cooper pair concentration. The constraint that the form factor $f$ imposes is evident: the measurement device must be properly calibrated to account for the effects of its geometry on the accumulated phase. For a given geometry, the smaller the critical current $I_{\rm c}$, the better the system for small acceleration measurement assuming that noise effects are mitigated. For instance, if $I_{\rm c}$ is in the $\mu$A range or below, the proposed system can detect small accelerations; it could thus be used with, e.g., disturbance reduction systems to accurately control the motion of a spacecraft driven by micronewton thrusters \cite{Morris2013,Levchenko2018}. Note that a deviation from linear response occurs if the ring is not perfectly symmetric. A difference $\Delta R_{\rm s}$ between the arms yields a quadratic contribution to the phase difference $m \pi R_s\Delta R_{\rm s} a^2/(2\hbar v^3)$. For currents close to $I_{\rm c}$, this effect is negligible if $a\Delta R_{\rm s} \ll (mR_{\rm s}^2 I_{\rm c} /(\hbar f ))^2$.

Turning to the measurement sensitivity, we extend first the SQUID model to account for possible loop inductance effects and see conditions to mitigate these. Assuming that the loop has inductance $L$, Eq.~\eqref{dphir} now reads: $\delta\phi= fa/I - L J/\Phi_0$, where $J$ is the current inside the loop \cite{SuppMat}. To minimize the SQUID surface, we consider an elongated rectangle: $b \ll c$, in which case $L\approx 2\mu_0(b-d)c/\pi d$. For the inductance to be negligible, say $L \sim 1 \rm{nH}$, the following condition must be fulfilled: $LI_{\rm c}/ \Phi_0 \ll 1$, which is in principle possible even for large $c$ because $d$ can be made close to $b$. Practically, for $d=10\mu m$ we can choose the ratio $(b-d)/d= 0.1$ and obtain $L \sim 1 \rm{nH}$ for $c=1$ cm. While for magnetometers the resolution is assessed against the flux noise $S_{\Phi}^{1/2}$, here we consider the sensitivity to acceleration $S^{1/2}_{a}$. \cite{SuppMat} We assume that the Josephson junctions are shunted with a resistance $R$ to cancel any capacitance effect. The voltage noise is $S_{\rm V} = 4k_{\rm B}TR$ at temperature $T$, and the transfer function is $\partial V/\partial \Phi = \pi RI_{\rm c}/\Phi_0$. \cite{SuppMat} Calculations yield: $S^{1/2}_a = 4/(Rf) S^{1/2}_V$. With $c=1$ cm, $f \simeq 5\times 10^{-2}$ A/(m$\cdot$s$^{-2}$); so with $T_{\rm c}=1.2$ K and a shunt resistance $R=2~\Omega$, we obtain $S^{1/2}_a= 4.6\times10^{-10} {\rm m\cdot s^{-2}/\sqrt{Hz}}$. This illustrative case shows promise for high resolution measurements and, crucially, demonstrates that the relevant geometrical factor is not the loop surface, but the form factor $f$, which implies possible minimization of the loop surface to mitigate magnetic flux noise effects.

The phase difference obtained so far relies on the validity of Eqs.~\eqref{eq:phi} and \eqref{eq:phi2}, which assume that the transverse acceleration to which the SQUID is subjected affects the phase but does not influence the Cooper pairs' trajectory. In reality the Copper pairs may well modify their phase due to their peculiar dynamics inside the superconductors involving a deviation of the actual center of mass motion trajectory $\mathbf{{r}}_{\pm}(t)$ of the superconducting fluid from the ideal one $\mathbf{{r}}_{0\pm}(t)$ centered along the wires direction. Here, to evaluate the magnitude of this possible deviation and check the reliability of our proposed SQUID-based accelerometer scheme, we use a more refined model. A Bose-condensed system evolves according to the Gross-Pitaevskii (GP) equation in the static frame \cite{annett2001james, pitaevskii2016bose}, which in our case results in two equations for each interferometer arm:
 \begin{eqnarray}
 \nonumber
&&i\hbar \partial_{t} \Psi_{\pm}(\mathbf{r},t) =
    \\
    &&\left[-\frac{\hbar^{2}}{2m} \partial^{2}_{\mathbf{r}}+\!V_\pm({\bf r}-{\bf r}_{0\pm}(t),t)\!+\!\rm{g_c} |\Psi_{\pm}(\mathbf{r},t)|^{2}\right]\!\!\Psi_{\pm}(\mathbf{r},t)~~~ 
\label{eq:b2p}
\end{eqnarray}
where $V_\pm({\bf r}-{\bf r}_{0\pm}(t),t)$ is the confining potential centered at the trajectories ${\bf r}_{0\pm}(t)$ going from one arm end to the other, and $\rm{g_c}$ is the low-energy scattering-length-dependent coupling constant characterizing the boson-boson interaction. Using the following transformations
\begin{equation}
\Psi_\pm({\bf r},t)= e^{i[{\bf k}_\pm(t)\cdot({\bf r}-{\bf r}_{\pm}(t))-\phi_\pm(t)]}\Psi_{0\pm}({\bf r}-{\bf r}_{\pm}(t),t)
\end{equation}
we rewrite Eq.~\eqref{eq:b2p} in the moving frame of the condensate center ${\bf r}_{\pm}(t)$ as:
\begin{eqnarray}
    &&i\hbar \partial_{t} \Psi_{0\pm}(\mathbf{r'},t)  =\left[ - \frac{\hbar^{2}}{2m} \partial^{2}_{\mathbf{r'}}  + g |\Psi_{0\pm}(\mathbf{r'},t)|^{2}\right]\Psi_{0\pm}(\mathbf{r'},t) \nonumber \\
    &+& \sum_{i=2}^{\infty} \frac{ \partial^{i} V(\mathbf{r}_{\pm} - \mathbf{r}_{0\pm}, t) }{ \partial \mathbf{r}_{\pm}^{i}}\frac{(\mathbf{r'})^{i}}{i!} \Psi_{0 \pm}(\mathbf{r'},t), 
\label{eq:b3p}
\end{eqnarray}
which is the modified GP equation for the condensate wave function with $\mathbf{r}' = \mathbf{r} -\mathbf{r}_{\pm}(t)$, provided that the following conditions are fulfilled:
\begin{eqnarray}\label{pos}
{\bf \dot r}_{\pm}(t)&=&\hbar{\bf k}_{\pm}(t)/m 
 \\ \label{mom}
\hbar{\bf \dot {\bf k}}_{\pm}(t)&=&-\frac{\partial V({\bf r}_{\pm}(t)-{\bf r}_{0\pm}(t),t)}{\partial{\bf r}_{\pm}(t)} 
\\
\phi_{\pm}(t)&=&-
\int_0^t \frac{dt'}{\hbar}L({\bf r}_{\pm}(t'),{\bf \dot r}_{\pm}(t'),t')
\\ 
\nonumber
\mbox{with}~~~~~~~~~~~~~ &&\\ \label{lag}
L_\pm({\bf r}_{\pm}(t),{\bf \dot r}_{\pm}(t),t)
&=&
\frac{m}{2}{\bf \dot r}^2_{\pm}(t) -
V({\bf r}_{\pm}(t)-{\bf r}_{0\pm}(t),t)
\nonumber
\end{eqnarray}
The above set of equations constitutes a general result which has a large range of applicability for interferometry and hence it fits our purposes. In atom interferometry, it provides a general expression for the phase evolution of an atomic cloud bucket with a parabolic potential confinement \cite{navez2016matter}. In the case of a SQUID however, the confining potential has the shape of a wall delimited by the superconducting wires and the cloud consists in a superconducting fraction of infinite size along the wires. Under the assumption that the center of mass motion follows exactly the center of the wires i.e. ${\bf r}_{\pm}(t)={\bf r}_{0\pm}(t)$ and denoting $\mu$ the chemical potential of the boson system, the transformed wave function $\Psi_{0\pm}(\mathbf{r},t)=\exp(-i\mu t/\hbar )\Psi_{0\pm}(\mathbf{r})$ then satisfies the static GP equation (\ref{eq:b2p}) with ${\bf r}_{0\pm}(t)=0$, hence proving Eqs.~\eqref{eq:phi} and \eqref{eq:phi2}. Both the chemical potential and the coupling constant of the GP equations can be estimated in terms of the superconductor's characteristics \cite{gen66}:
\begin{eqnarray}
\mu=\frac{\hbar^2 (T_{\rm c}-T)}{2m\xi_0^2 T_{\rm c}} \quad 
{\rm g_c}=0.107\left(\frac{\hbar^2}{2m\xi_0^2}\right)^2\frac{N(0)}{k_{\rm B} T_{\rm c}}
\label{eq:bd}
\end{eqnarray}
where $\xi_0$ is the coherence length, $T_{\rm c}$ the critical temperature, and $N(0)=m^{2}v_{\rm F}/(2 \pi^{2} \hbar^{3})$ the density of state at the Fermi surface with $v_{\rm F}$ being the Fermi velocity  \cite{gen66}. Any deviation from the ideal center of mass motion trajectory resulting in the phase perturbation of the wave function $\Psi_{0\pm}(\mathbf{r},t)$, can now be estimated. Time-dependent perturbation theory applied to Eq.~(\ref{eq:b3p}) \cite{landau2013quantum} and translation invariance along the wire, yield the difference between the real trajectory and the ideal one, which has only a transverse component. Estimated for the ring, the difference should be much smaller than the wavelength of the macroscopic wave function \cite{SuppMat}: 
\begin{equation}
|{\bf r}_{\pm}(t)-{\bf r}_{0\pm}(t)|\simeq 
\frac{md^2 a}{16\mu} \sim \frac{m^2d^2\xi_0^2a}{8\hbar^2}\ll \frac{R_{\rm s}}{\phi_\pm}\sim \frac{\hbar}{mv}  
\end{equation} 
For the ring geometry in the ``worst-case scenario'', the radius $R_{\rm s}$ may increase or decrease by the width $d$ of the wire: $R_{\rm s} \pm d$. Therefore, for a thin wire this change does not affect much the phase difference for small winding numbers in the phase accumulation for each arm; we checked that the condition $\phi_\pm/\pi = 2m^2R_{\rm s}^3I/(\hbar^2 f)\leq 1$ is fulfilled.\\

We theoretically showed how existing superconducting devices can be employed as inertial sensors for determining the acceleration by measuring the current or the voltage. We propose that SQUIDs are not used as magnetometers for acceleration measurements but directly as interferometric accelerometers. Crucially for that purpose, using a time-dependent Gross-Pitaevskii model, we proved the Cooper pairs' phase sensitivity to inertia. Note that the Cooper pairs transport is stable against local temperature gradients induced by thermal fluctuations since their thermoelectric coupling is zero. We also checked the potential high resolution of a SQUID used as an accelerometer. We found that external magnetic flux noise is not a problem as $S_{a}^{1/2}$ depends on the form factor $f$, which implies that one can reduce the loop surface. We suggest that the effectiveness (portability, cost, accuracy) of SQUID-based interferometric accelerometers is experimentally tested as geometrical parameters, superconducting materials, critical current values and circuit design offer a rich field of play for the system's optimization. We finally emphasize that our suggested approach may pave the way for further investigations of the Sagnac effect, or gravitational waves detection up to the high-frequency band.

\section*{Supplementary Material} Mathematical details omitted in the main text are given in the Supplementary Material below where a simple model of rf SQUIDs for vibration measurements is also discussed.

\begin{acknowledgments}
The authors thank Dr. Giampiero Marchegiani and Dr. Boris Chesca for discussions. PN acknowledges support by the EC Horizon 2020 Framework Programme project SUPERGALAX (Grant agreement ID: 863313).
\end{acknowledgments}

\newpage

\begin{widetext}

\appendix

\subsection{Phase accumulation for the rectangular geometry}
\label{AppA}

Here, we consider the rectangular geometry as shown in Fig.~\ref{fig:a1} and a transverse acceleration along the $y$ axis. 

\begin{figure}[h]
  \includegraphics[angle=0,origin=c, width=.5\textwidth]{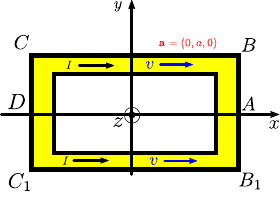}
  \vspace*{-0.4cm}
  \caption{Representation of the SQUID with a rectangular geometry.} 
  \label{fig:a1}
\end{figure}

The accumulated phase can be calculated as follows \cite{navez2016matter}:
\begin{equation}
\label{eq:a1}
\phi = \int_{t_{1}}^{t_{2}} \frac{m\mathbf{\dot{r}}_{\pm}^{2}}{2\hbar} dt, 
\end{equation}
where $\mathbf{r}_{\pm}(t)$ is the radius vector at time $t$. Denoting $\mathbf{v}$ the Cooper pair velocity relatively to the device, we have along the segment $AB$, $\mathbf{r}_{+}(t)= (0, -v_{y}t+at^{2}/2, 0)$, $\mathbf{\dot{r}}_{+}(t)= (0, -v_{y}+at, 0)$, and $t \in [0, t_{1}]$, where $t_{1}=c/{v_{y}}$. For the lower part, $AB_{1}$, the corresponding parameters are $\mathbf{r}_{0-}(t)= (0, v_{y}t+a_{y}t^{2}/2, 0)$, $\mathbf{\dot{r}}_{-}(t)= (0, v_{y}+at, 0)$, and $t \in [0, t_{1}]$, where $t_{1}=c/{v_{y}}$. The phase difference $\delta \phi_i =\phi_{AB}-\phi_{AB_{1}}$ is given by 

\begin{equation}
\label{eq:a2}
\delta \phi_{i} = \frac{m}{2\hbar} \int_{0}^{t_{1}} \left [ ((-v_{y}+at)^{2}-(v_{y}+at)^{2} \right] {\rm d}t = -\frac{mv_{y}a}{\hbar} t_{1}^{2}
\end{equation}

The next step is the calculation of the phase difference for the lower and upper parts of the junction. On the line $BC$, we have $\mathbf{r}_{0+}(t)= (v_{x}t, at^{2}/2, 0)$, $\mathbf{\dot{r}}_{0+}(t)= (v_{x}, at, 0)$ and  $\mathbf{r}_{0-}(t)= (v_{x}t, at^{2}/2, 0)$, $\mathbf{\dot{r}}_{0-}(t)= (v_{x}, at, 0)$ on the line $B_{1}C_{1}$. Here, $t \in [ t_{1}, t_{2}]$, where $ t_{2}= 2b/v_{x} +t_{1}$. As we assume an equal current $I$ for each side of the junctions, which is related to velocity as $I= envS$, so that $v_{x}=v_{y}=v$. Thus, the phase difference $\delta \phi_{ii}=\phi_{BC}-\phi_{B_{1}C_{1}}$ is
\begin{equation}
\label{eq:a3}
\begin{aligned}
\delta \phi_{ii} &= \frac{m}{2\hbar} \int_{t_{1}}^{t_{2}} \left [ (v_{x})^{2}+(at)^{2}-(at)^{2}-(v_{x})^{2} \right] {\rm d}t =0 . 
\end{aligned}
\end{equation}

Last, we consider the upper part $CD$, with $\mathbf{r}_{0+}(t)= (0, v_{y}t+at^{2}/2, 0)$ and $\mathbf{\dot{r}}_{0+}(t)= (0, v_{y}+at, 0)$, and the lower part $C_{1}D_{1}$, with $\mathbf{r}_{0-}(t)= (0, -v_{y}t+at^{2}/2, 0)$ and $\mathbf{\dot{r}}_{0-}(t)= (0, -v_{y}+at, 0)$. The integration in the range $t \in [ t_{2}, t_{3}]$, where $t_{3}=c/v_{y}+t_{2}=t_{1}+t_{2}$, yields the phase difference $\delta \phi_{iii}$:
\begin{equation}
\label{eq:a4}
\delta \phi_{iii} = \frac{m}{2\hbar} \int_{t_{2}}^{t_{3}} \left [ (v_{y}+at)^{2}-(-v_{y}+at)^{2} \right] {\rm d}t = \frac{mv_{y}a}{\hbar} (t_{3}^{2}-t_{2}^{2}). 
\end{equation}

The summation of phases $\delta \phi=\delta \phi_{i}+\delta \phi_{ii}+\delta \phi_{iii}$ leads to the final result  
\begin{equation}
\label{eq:a5}
\begin{aligned}
\delta \phi = \frac{mv_{y}a}{\hbar} (t_{3}^{2}-t_{1}^{2}-t_{2}^{2}) = 
\frac{2c(2b+c)}{v} \frac{ma}{\hbar}
\end{aligned}
\end{equation}
taking $v_x=v_y=v$.

\subsection{Effect of acceleration on a Cooper pair trajectory}
\label{AppB}
We seek to find the deviation $\delta \mathbf{r}$ of the center of mass of the system due to the acceleration to which the condensate is subjected. Unlike situations where thermal electrons localized on the surface may screen an external force field inside the bulk which, therefore, may not be effective \cite{Tolman1913,Tolman1916,Fischer2001}, for temperatures close to the absolute zero, the acceleration field is effectively present inside the superconductor bulk. To establish an upper bound for the trajectory deviation ${\bf r}_{\pm}(t)-{\bf r}_{0\pm}(t)$, we consider the worst case scenario of an acceleration perpendicular to the wire, assumed here to be infinitely long and centered at ${\bf{r}}_0=0$. In this situation, the Cooper pairs can be described with the Gross-Pitaevskii equation: 
 \begin{equation}
 \begin{aligned}
    i\hbar \partial_{t} \Psi(\mathbf{r},t)=
    \left(-\frac{\hbar^{2}}{2m} \partial^{2}_{\mathbf{r}}  + V({\bf r},t) + m\mathbf{a}.{\bf r} + {\rm g} |\Psi_{\pm}(\mathbf{r},t)|^{2} \right)\Psi_{\pm}(\mathbf{r},t),
\end{aligned}
\label{eq:g1}
\end{equation}
where $\mathbf{a} = (0,0,-a)$. 

In the Thomas-Fermi limit, the wave-function of the condensate is given by  
 \begin{equation}
 \begin{aligned}
    \Psi(\mathbf{r},t)= \mathrm{exp} \bigg ( \frac{-i\mu t}{\hbar}\bigg) \sqrt{\frac{\mu}{{\rm g}}}\sqrt{1 - \frac{V({\bf r},t)}{\mu } - \frac{m\mathbf{a}.{\bf r}}{\mu }},
\end{aligned}
\label{eq:g2}
\end{equation}
with $\mu$ being the chemical potential of the condensate. The density of the condensate $n(\mathbf{r},t)$ then reads

 \begin{equation}
 \begin{aligned}
    n(\mathbf{r},t) = |\Psi(\mathbf{r},t)|^{2} = \frac{\mu}{{\rm g}}\left(1 - \frac{V({\bf r},t)}{\mu } - \frac{m\mathbf{a}.{\bf r}}{\mu } \right)
\end{aligned}
\label{eq:g3}
\end{equation}

\noindent For simplicity, we consider a wire potential 
\begin{equation}
V(\mathbf{r},t) =
    \begin{cases}
      0 & \text{if  $ r=\sqrt{r_y^2+r_z^2} \leq d/2 $ }\\
      \infty & \text{otherwise}
    \end{cases}
\label{eq:g5}
\end{equation}
with $d$ being the diameter of the wire section. 
The deviation reads:
\begin{equation}
 \begin{aligned}
    \delta \mathbf{r} = \frac{\displaystyle \int_{0}^{\infty}\!\!\!\int_0^{2\pi} n(\mathbf{r},t) \mathbf{r} r d r d\theta }{\displaystyle \int_{0}^{\infty}\!\!\!\int_0^{2\pi} n(\mathbf{r},t) r d r d\theta}
\end{aligned}
\label{eq:g4}
\end{equation}
and as only the $z$-component of the center mass deviation due to the acceleration field is nonzero, we find:
\begin{equation}
\delta r_{z} = \frac{\displaystyle \frac{\displaystyle \mu}{g}  \int_{0}^{d/2}\!\!\!\int_0^{2\pi} r\sin \theta \left( 1 + \frac{m a r \sin \theta}{\mu}\right) r{\rm d}r {\rm d}\theta}
{\displaystyle \frac{\mu}{g} \int_{0}^{d/2}\!\!\!\int_0^{2\pi} \left( 1 + \frac{m a r \sin \theta}{ \mu}\right)  r{\rm d}r {\rm d}\theta} = \frac{mad^{2}}{16\mu}
\label{eq:g6}
\end{equation}

We may now give an order of magnitude for the deviation using $\mu=\hbar^2 (T_{\rm c}-T)/(2m\xi_0^2 T_{\rm c})$ and the following figures that may describe a real-life setting: the coherence length is $\xi_{0}=100$ nm \cite{annett2001james}, $d = 10 \mu$m, and $a = 0.1$m$\cdot$s$^{-2}$. We thus obtain:  
\begin{equation}
 \delta r_{z} = \frac{m^{2}d^{2} \xi_{0}^{2}}{8 \hbar^{2}}a \bigg( 1 -\frac{T}{T_{\rm c}} \bigg )^{-1} \geq \frac{m^{2}d^{2} \xi_{0}^{2}}{8 \hbar^{2}}a \approx 3.7 \times 10^{-18} \mathrm{m}.
\label{eq:g9}
\end{equation}
which is a negligible for practical applications.

\subsection{Oscillatory motion measurement}
As vibrations whether free, forced or damped, are ubiquitous, accurately sensing and measuring vibratory phenomena is of great practical importance \cite{Vibrations2017}. Vibrations in the nanometer range and up to megahertz can either be a desired phenomenon or an unwanted effect, which may act like noise during measurement. Managing to characterize noise permits to identify its contribution to the accumulated phase. So, we also consider oscillatory motion in our work.

As we now turn to vibrations the core component of the accelerometer is a radio frequency (rf) SQUID based on the ac Josephson effect, as illustrated in Fig.~\ref{circuit} and discussed in Ref.~\cite{schwartz2013superconductor}. Usually in a rf scheme, a loop with a Josephson junction is inductively coupled to a tank circuit. We denote $I_-$ the current that goes through the inductance $L$ of the loop, and $I_+$ the ac current with frequency $\omega$ that goes through the Josephson junction. We impose a dc current $I_{\rm dc}$ through $L$ as a working condition so that $I_-=I_{\rm dc}$, which permits acceleration measurements without having to deal with an additional phase. In absence of acceleration, the phase accumulation is equal in the two branches of the loop: $\arcsin(I_+(t)/I_{\rm c})+ \phi_+=\phi_-$ and imposes also dc current in addition to $I_+$ through the Josephson junction, but this dc contribution is negligible as long as $m^2R_{\rm s}^3 I_{\rm c}\ll \hbar^2 f$.

The voltage across the rf SQUID is related to the change of the electromotive force acting on the Cooper pairs caused by the acceleration which induces a phase difference:
\begin{eqnarray}
V= - \frac{\Phi_{0}}{2\pi}~\partial_t\delta \phi,
\label{eqvolt}
\end{eqnarray}
In other words, the term $\partial_t\delta\phi$ is related to the kinetic inductance or Josephson inductance $L_{\rm J}=\Phi_0/(2\pi I_{\rm c})$. For oscillating accelerations of the type: $a(t)=a_{\omega} e^{i\omega t}+ c.c.$, with $a_{\omega}$ being the Fourier component of the acceleration in the frequency space, we write the alternating Cooper pairs current as $I_+(t)$ and $I_-(t)=I_{\rm dc}- I_+(t)$ with a magnitude $|I_{\omega}| \leq I_{\rm c}$. The amplitude of the current $I_{\omega}$ is subsequently measured using the tank.

\begin{figure}[h]
  \includegraphics[angle=0,origin=c,width=.4\textwidth]{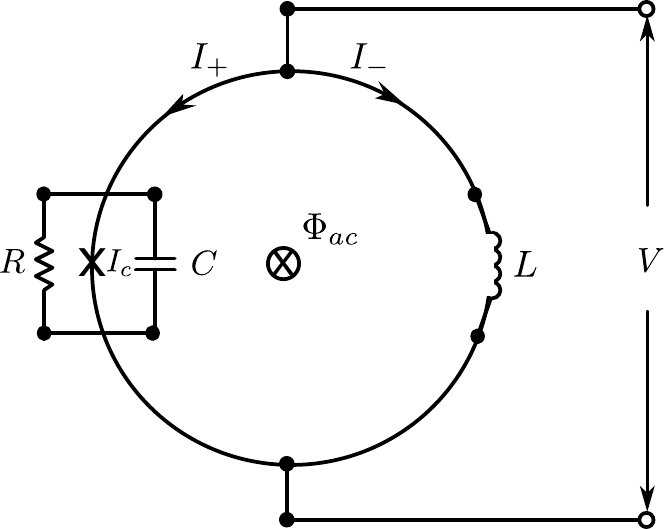}
  \caption{Circuit model of the rf SQUID-based accelerometer. The Josephson junction with inductance $L_{\rm J}$, is represented by a cross; it is shunted by the resistance $R$ and the capacitance $C$. The superconducting ring has an inductance $L$.
  }
  \label{circuit}
\end{figure}

Assuming that the Cooper pairs' drift velocity $v$ is such that the acceleration is constant over the travelling time: $R/v \ll 2\pi/\omega$, the total flux reads $\delta \phi(t)\Phi_{0}/2\pi = -LI_-(t) + \Phi_{\rm ac}(t)$, with $\delta\phi$ satisfying \cite{barone1982physics}: 

\begin{equation}
\left(LC\partial_t^2  + \frac{L}{R} \partial_t + 1\right)\delta \phi + \frac{2\pi LI_{\rm c}}{\Phi_{0}}  \sin (\delta \phi) = \frac{2\pi \Phi_{\rm ac}}{\Phi_{0}}
\label{kirg}
\end{equation}

\noindent and $\Phi_{\rm ac}(t) = (2 \pi)^{-1}\Phi_{0} fae^{i\omega t}/I_{\rm dc} +c.c.$ being the effective magnetic flux due to the acceleration associated with the oscillatory motion. For weak current changes, $I\approx 2I_{c} \delta \phi$, Eq.~\eqref{kirg} becomes:
\begin{eqnarray}
\left(\partial_t^2  + \frac{1}{RC} \partial_t\right)\delta \phi + \frac{1}{L_{\rm J}C}\delta \phi = - \frac{2\pi I_-}{C\Phi_{0}}
\label{eq:o3b}
\end{eqnarray}
\noindent This equation has the form of an equation of motion for a driven harmonic oscillator:
\begin{eqnarray}
\ddot{x} + 2\zeta \omega_{0} \dot{x} + \omega_{0}^{2} x = -\frac{1}{L_{\rm J}I_{\rm c}C}I_{\omega}e^{i\omega t} +c.c.,
\label{eq:o4}
\end{eqnarray}
where $x=\delta\phi$, $\omega^{2}_{0} = 1/(L_{\rm J}C)$, and $\zeta = 1/(2RC\omega_{0})$ is the damping rate. So, from its solution we obtain the following expressions for $\Phi_{\rm ac}(t)$ and $V(t)$:

\begin{equation}
\label{eq:o6}
\Phi_{\rm ac}(t) = \Phi_{\omega}e^{i\omega t} +c.c. ~ \mbox{with}~ \Phi_\omega= \!\frac{-iI_{\omega}}{C(\omega^{2}_{0} + 2i \zeta\omega\omega_{0}  - \omega^{2})}
\end{equation}
\begin{equation}
V(t) = V_\omega e^{i\omega t}+c.c. ~ \mbox{with}~ V_{\omega} = \!\frac{iI_{\omega}\omega}{C(\omega^{2}_{0} + 2i \zeta\omega\omega_{0}  - \omega^{2})}
\label{eq:o7}
\end{equation}

Introducing the frequency $\omega_{0} = 1/\sqrt{L_{\rm J}C}$ and the damping factor $\zeta = 1/(2RC\omega_{0})$, and inserting Eq.~\eqref{eqvolt} into Eq.~\eqref{kirg}, yield the time-dependent voltage $V(t)=V_\omega e^{i\omega t}+c.c.$ with
\begin{equation}
V_\omega= \frac{iI_{\omega}\omega}{C(\omega^{2}_{0} + 2i \zeta\omega\omega_{0}  - \omega^{2})}
\label{eq:fv}
\end{equation}
Finally, we may relate the magnitude of the acceleration $a(t)$ to the alternating Cooper pairs current and the voltage amplitudes $I_{\omega}$ and $V_{\omega}$:

\begin{eqnarray}
a_\omega & = & \frac{2\pi I_{\rm dc}}{f\Phi_{0} }
\left(L + \frac{1}{ C(\omega^{2}_{0} + 2i \zeta\omega\omega_{0}  - \omega^{2})}\right)I_{\omega}\\
& = & \frac{2\pi I_{\rm dc}}{f\Phi_{0}}
\left(\frac{L}{Z_{\omega}} + \frac{1}{i \omega}\right)V_{\omega}
\label{eqaccel}
\end{eqnarray}
 
\noindent with $Z_{\omega}=V_{\omega}/I_{\omega}$ being the circuit's the impedance. Note that in the limit $\omega \xrightarrow{} 0, L \xrightarrow{} 0$, we recover the expression for the transverse acceleration $\lim_{\omega, L \to 0} a_{\omega} =  I_{\rm dc}/f\Phi_{0}$.

For a given fixed frequency $\omega$, $|I_\omega|$ and hence $|V_{\omega}|$ grow linearly with $a_{\omega}$. The frequency range achievable for acceleration measurement is bounded from above by the passage time $\pi/\Omega$ of the Copper pair through the ring, so the condition $\omega \ll \Omega= 4mR_{\rm s} I_{\rm dc}/(\hbar f)$ or, equivalently, $\omega \ll \hbar I_{\rm dc}/(mR_{\rm s}^2 I_{\rm c})$, must be satisfied. For, e.g., $R_{\rm s}=10~\mu$m this condition limits the frequency to the $\rm MHz$ range. Note that in practice, the variation of flux can be related the measured voltage $V$. Figure~\ref{fig:4} shows the imaginary and the real parts of the voltage $V_{\omega}$ against the frequency $\omega$ for a simple model system subjected to an acceleration amplitude $|a_\omega|$ of $10^3$ m$\cdot$s$^{-2}$. Note the high-frequency resonance at several MHz, which allows for high $g$-force acceleration measurements. One  possible application of our proposed SQUID-based accelerometer could thus be the detection of ultrasonic vibration in nanosystems up the MHz range regime. A practical example is that of tool state monitoring to prevent wear and breakage \cite{thomas2000silicon}.

\begin{figure}[h]
  \centering
  \includegraphics[width=.5\textwidth]{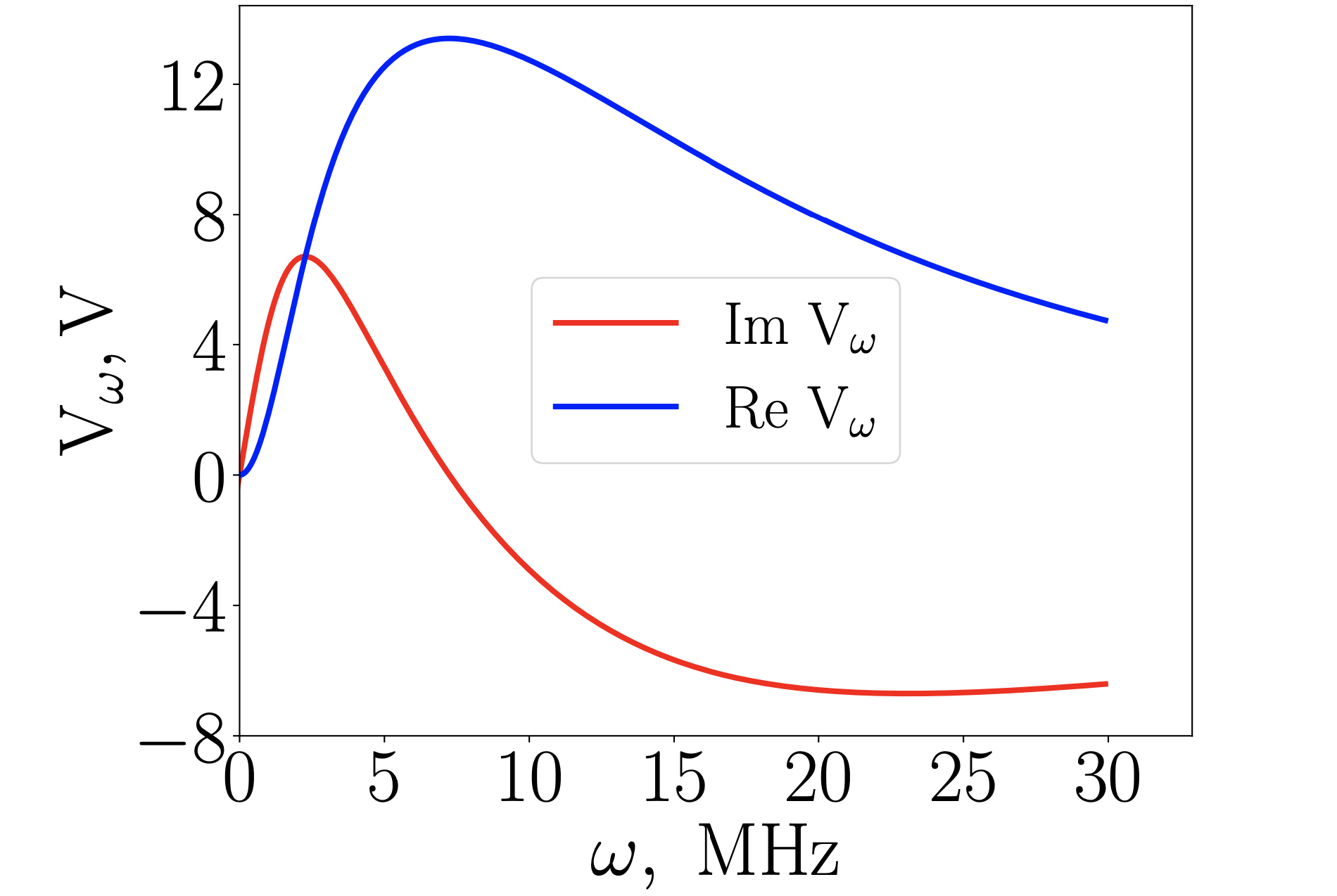}
  \caption{Imaginary and real parts of the voltage $V_{\omega}$ as a function of the frequency $\omega$.
  The curves are shown for an  acceleration $a_{\omega}=10^{3}$ m$\cdot$s$^{-2}$, a Josephson inductance $L_{\rm J}=0.66$ nH, a capacitance $C=4.8 \cdot 10^{-5}$ F (hence $\omega_{0} = 5.62$ MHz), an external inductance $L=1$ nH, and a resistance $R= 1 \mathrm{m}\Omega$.} 
  \vspace*{-0.4cm}
  \label{fig:4}
\end{figure}

\subsection{Sensitivity to acceleration} 
In experiments, the voltage noise, $S^{1/2}_{V}$, and the flux noise, $S^{1/2}_{\Phi}$, are measured to quantify the sensitivity of the SQUID. Assuming that each Josephson junction is shunted with a resistance $R$, the voltage noise originates in the Johnson noise so that $S_{V} = 4 k_{\rm B}TR$ \cite{Tesche1977,Tesche1979}. The flux noise $S^{1/2}_{\Phi}$ is related to the voltage noise $S_{V}$ as follows: $S^{1/2}_{\Phi} = S^{1/2}_{V} / \partial V / \partial \Phi$, where $\partial V / \partial \Phi$ is the transfer function. 

To calculate $S^{1/2}_\Phi$, we need to derive an expression for $\partial V / \partial \Phi$ but first, we derive the time-averaged voltage from the solution of the equation that the total current satisfies:

\begin{equation}
I=2I_{\rm c}\sin\left(\frac{\phi_+ +\phi_-}{2}\right)\cos\left(\frac{\phi_+ -\phi_-}{2}\right)+\frac{2\Phi_0}{2\pi R}\frac{{\rm d}(\phi_+ +\phi_-)}{{\rm d}t} 
\end{equation}

\noindent Note that here, as in Ref.~\cite{Oppenlander2000}, we can neglect capacitive effects in the junctions provided that the shunt resistance used is small enough. Imposing $I$ as a constant current bias, the time-averaged voltage reads:

\begin{equation}
V=R\sqrt{(I/2)^2 -I_{\rm c}^2\cos^2[(\phi_+-\phi_-)/2]}
\end{equation}

\noindent Next, using the definition of the phase difference $\delta \phi$, we get
\begin{equation}
\frac{\partial}{\partial \Phi} = \frac{2 \pi}{ \Phi_{0}} \frac{\partial}{\partial  \delta \phi}
\end{equation}

\noindent and the transfer function reads:

\begin{equation}
\frac{\partial V}{\partial \Phi} = \frac{2 \pi}{ \Phi_{0}} \frac{R I^{2}_{\rm c}}{4} \frac{\sin(\phi_+-\phi_-)}{\sqrt{(I/2)^{2}-I^{2}_{\rm c} \cos^{2}[(\phi_+-\phi_-)/2]}}.  
\end{equation}

\noindent Choosing $I=2I_{\rm c}$, which is the condition for the onset of the voltage, we now obtain:

\begin{equation}
\frac{\partial V}{\partial \Phi} \approx \frac{2 \pi}{\Phi_{0}} \frac{R I^{2}_{\rm c}}{2 I_{\rm c}} = \pi \frac{R I_{\rm c}}{\Phi_{0}} 
\end{equation}

\noindent and we may now compute the flux noise using some particular values of the model parameters for two cases. We assume the following geometries: $R_{\rm s} = b = c = 300 \mu$m. Case 1: $T= 1.2$ K, $I_{\rm c} = 0.5~\mu$A, and a shunt resistance of $R=1$ m$\Omega$; Case 2: $T= 1.2$ K, $I_{\rm c} = 0.5~\mu$A, and a larger shunt resistance of $R=2~\Omega$ that still cancels the effect of the capacitance. Numerical results are as follows:
 
\begin{center}Case 1: $S^{1/2}_{\Phi} \approx 164 \mu\Phi_{0}/\sqrt{\mathrm{Hz}}$.\\ Case 2: $S^{1/2}_{\Phi} \approx 3600 n\Phi_{0}/\sqrt{\mathrm{Hz}}$.
\end{center}

The latter figure is an order of magnitude larger than state-of-the-art reference. We thus see that if used as a magnetometer the SQUID does not boast a performance level on par with state-of-the-art resolution. However, methods exist to reduce effects of the background noise and improve the signal-to-noise ratio \cite{He2003}. But more importantly, our proposal is to \emph{not} use the SQUID as a magnetometer but as an accelerometer.

We must now derive the SQUID sensitivity to acceleration $S_{a}^{1/2}$ and see if and how the geometry matters. For that purpose, we even extend the model to account for possible effects of an inductance and see conditions to mitigate these. The equation for the phase difference reads: 
\begin{equation}
\phi_+ -\phi_-= \frac{fa}{I} - L J/\Phi_0
\end{equation}

\noindent where $J$ is the current inside the loop given by:

\begin{equation}
J=I_{\rm c}(\sin \phi_+ -\sin \phi_-)+ \frac{\Phi_0}{2\pi R}\left(\frac{{\rm d}\phi_+}{{\rm d}t} -\frac{{\rm d} \phi_-}{{\rm d}t}\right)
\end{equation}
Considering an elongated rectangle $b \ll c$ to minimize the SQUID surface, we obtain the following expression for the inductance:

\begin{equation}
L=2\frac{\mu_0 c}{2\pi} \ln\left(\frac{b-d/2}{d/2}\right)
\stackrel{b \rightarrow d}{=} 4\frac{\mu_0 c}{2\pi} \frac{(b-d)}{d}
\end{equation}

For the inductance to be negligible for the stationary case, the following condition must be fulfilled:
$LI_{\rm c}/ \Phi_0 \ll 1$, which is in principle possible even for large $c$ because $d$ can be made close to $b$. Practically, for $d=10\mu m$ we can choose the ratio $(b-d)/d= 0.1$ and obtain $L \sim 1 \rm{nH}$ for $c=1$ cm. This value allows to estimate the sensitivity to the acceleration $S^{1/2}_{a}$ as follows. 

From $I=2I_{\rm c}$, the voltage generated as the SQUID undergoes a small acceleration, which using Eq.~(3) of the manuscript, can be approximated by:

\begin{equation}
V\simeq RI_{\rm c}(\phi_+-\phi_-)/2= R f a/4
\end{equation}

\noindent where $f$ is the form factor defined in Eq.~(4) of our manuscript. We then deduce:

\begin{equation}
S^{1/2}_a = 4\pi I_{\rm c}/(f \Phi_0) S^{1/2}_\Phi= 4/(Rf) S^{1/2}_V
\end{equation}

\noindent With $c=1$ cm, $f \simeq 5\times 10^{-2}$ A/(m$\cdot$s$^{-2}$); so with $T_{\rm c}=1.2$ K and $R=2~\Omega$, we obtain:

\begin{equation}
S^{1/2}_a= 4.6\times10^{-10} {\rm m\cdot s^{-2}/\sqrt{Hz}} = 4.6\times10^{-11} {g/\sqrt{\rm Hz}}
\end{equation}

\noindent These figures are a clear indicator that a practical implementation of our proposal could yield extremely high resolution performance, i.e. better than what atomic interferometry can achieve $\sim 10^{-8}$ $g/\sqrt{\rm Hz}$ \cite{LeGouet2008}, which can also be promising in experiments where gravity gradients affects the phase shifts \cite{DAmico2017}.

The magnetic sensitivity of a SQUID scales with its dimensions and this is an important aspect when a SQUID is used as a magnetometer. But when used as an accelerometer for direct measurement, the illustrative calculation above shows that a very high sensitivity to acceleration can be obtained. The key point here is that the sensitivity to acceleration does not depend on the SQUID's surface but rather on the form factor $f$, Eq.~(4) in the main text. This clearly opens up a new avenue for the design and testing of devices with a quasi-linear geometry, i.e. with the width $b \longrightarrow 0$ for which surface-dependent magnetic sensitivity plays no role.

\subsection{Discussion of possible side effects}

\subsubsection{General considerations}
Coulomb screening as well as the role played by the lattice ions which may distort the field acting on the electrons, can hinder the direct measurement of gravitation by SQUID interferometry, as these side effects result in an effective force that balances gravity. Few experimental evidence have been gathered since the very early works addressing the problem of gravitational effect on conduction electron is solid-state systems~\cite{Tolman1913,Tolman1916}. Among the  most recent papers, only two experimental works have been performed to analyze the effect of gravity on electrons~\cite{Witteborn1967, Witteborn1968}. Within 10\%, they recover the gravitation acceleration $g=9.81$ m$\cdot$s$^{-2}$. It is worth noticing that no subsequent experiment has confirmed this measured effect, and that the theoretical interpretation of this experiment was deemed  controversial~\cite{Schiff1970}. Now, more than 50 years after, the issue is not quite clear concerning the acceleration of electrons in metals, and we note that these works have not been further pursued.

It is also worth emphasizing that the experimental works were not done using superconductors but mostly normal metals. So, this does not tell whether screening effects occur in a superconductor given that the Cooper pairs have some coherence and penetration lengths. We have indeed stated that the normal electrons could screen the gravitational force, but to the best of our knowledge no one established up to now if the screening will occur with Cooper pairing. While one can speculate on theory-based assumptions,
only adequate experiments could answer on these issues. Since the SQUID is not a complicated structure and easy to use, it is also easier to do experiments rather than elaborate a sophisticated theory that might appear controversial, given also the fact that no theory exists for high-$T_{\rm c}$ superconductors leading still to many controversies.  

Here, we emphasize the two following points: First, the Landau Ginzburg model does not contain an explicit nonlocal Coulomb interaction term leading to screening and important variations of the electron charge density within a superconductor (namely with vortices). As it is a phenomenological approach, no clear explanation is provided but it could be that of a large coherence length or of a lattice deformation that could allow such density variations. Thus the problem of screening with superconductor somehow remains open.

2) The Gross-Pitaevskii model shows precisely that some internal deformation of the wave function is possible and that what matters is the center-of-mass motion across the two arms. Contrary to the case of a semi-classical system, the screening is not done by some electrons at the interface to the inner electrons, since the wave function is fully delocalized in the entire SQUID structure. So the wave function as a whole feels the acceleration despite its potential internal deformation due to screening.

\subsubsection{Influence of the screening effect: worst-case scenario}
In case of total screening, acceleration effects are negligible and no interference due to gravity is expected to be observed. But even in this worst-case scenario, we show that it is nevertheless possible to observe a phase difference provided that the relaxation time for transferring electrons between the two arms of the SQUIDS is greater than that to carry out a measurement. This requires four Josephson junctions and a larger capacitance between the arms.

In the new scheme presented in Fig.~\ref{figscreening} below, we assume that the gravitation force is exactly balanced by an electric force. This results in a charge distribution $Q_g$ at the two arms of the interferometers to compensate exactly the electric force $ma/2e$ induced by the acceleration. The electric potential is now $V_{12}=ma 2\overline{r}/ e$ where $\overline{r}$ is the average position  in the y-direction. For a surfaceless structure (rectangle with $b=0$), we  estimate by the structure symmetry that $\overline{r}= c/2$ and that the phase $\phi_\pm = \mp e V_g c /\hbar v$  as the average over time or position are equivalent. The potential created by the acceleration is related to the charge by a capacitance $V_{12}=(Q_g+Q)/C$. At the junctions between the two arms, a discharge transfers an amount $Q$ with the undesired effect to equalize the potential difference between the arms.

\begin{figure}
    \centering
    \includegraphics[width=.5\textwidth]{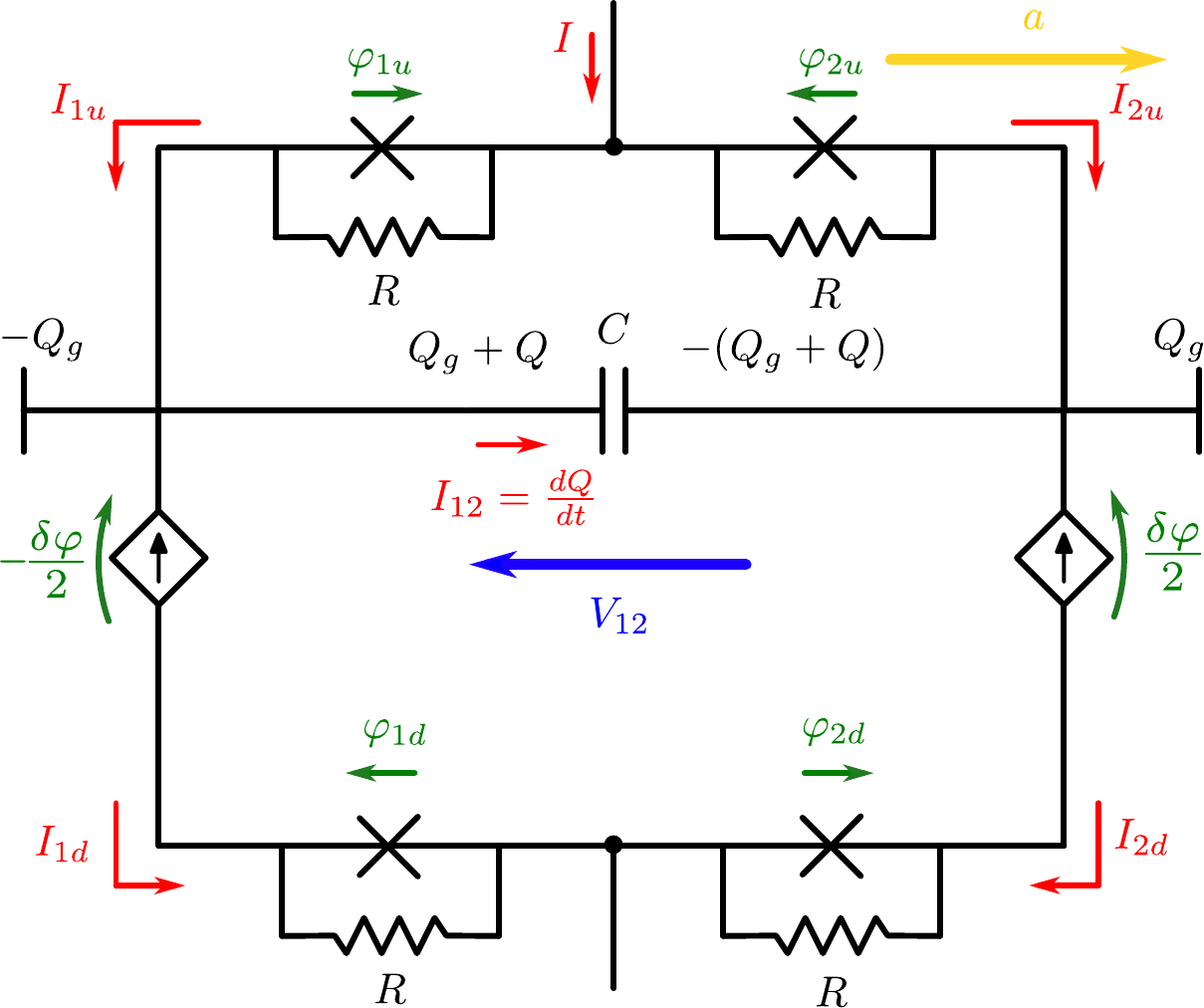}
    \caption{Scheme of the SQUID junction accounting for Coulomb screening effects.}
    \label{figscreening}
\end{figure}

Using the relations:
\begin{eqnarray}
I_{i,u/d}= I_{\rm c} \sin(\phi_{i,u/d})+ \frac{\hbar}{R 2e} \frac{\rm d}{{\rm d}t} \phi_{i,u/d}
\\
I_{12}=C\frac{\rm d}{{\rm d}t} V_{12} - \frac{\rm d}{{\rm d}t}Q_g
\end{eqnarray}
and the Kirchoff relations for the phases (or potential) and the currents, we get for $V=0$ (as convention the phase difference associated to $V$ is also set to zero) to the equations:
\begin{eqnarray}
\nonumber
&&C\frac{\rm d}{{\rm d}t}V_{12}-\frac{\rm d}{{\rm d}t}Q_g 
=-2I_{\rm c} \sin\left( \frac{\phi_{12}}{2}+\frac{\delta \phi}{4} \right)\cos(\frac{\delta \phi}{4})\\
&-&\frac{1}{R}\left(V_{12}+\frac{\hbar}{4e}\frac{\rm d}{{\rm d}t}\delta \phi  \right)
\\
&&\phi_{1u} +\phi_{1d}+\phi_-=\phi_{2u} +\phi_{2d}+\phi_+=0
\\
&&\phi_{1u}+\phi_{2u}=\phi_{1d} +\phi_{2d}
\\
I&=&2I_{\rm c} \cos\left(\frac{\phi_{12}}{2} +\frac{\delta \phi}{4}\right)\\
\nonumber
&\times&\sin\left(\frac{\phi_{1u} +\phi_{2u}+\phi_{1d} +\phi_{2d}}{4}\right)\cos\left(\frac{\delta \phi}{4}\right)
\end{eqnarray}
with $\delta\phi=\phi_+-\phi_-=\frac{f e}{2 m\overline{r}I_{\rm c}}V_{12}$. Linearizing the first equation, we deduce the maximum current:
\begin{eqnarray}
C_R \frac{{\rm d}^2}{{\rm d}t^2} \phi_{12}&+&
\frac{1}{R_R}\frac{\hbar}{2e}\frac{\rm d}{{\rm d}t}\phi_{12}+I_{\rm c} \phi_{12}=C \frac{2m\overline{r}}{e}\frac{\rm d}{{\rm d}t}a
\\
I_{\rm max}&=&2I_{\rm c} \cos\left(\frac{\phi_{12}}{2} +\frac{\delta \phi}{4}\right)
\cos\left(\frac{\delta \phi}{4}\right)
\\
\delta\phi&=&\frac{f e}{2m\overline{r}I_{\rm c}}\frac{\hbar}{2e}\frac{\rm d}{{\rm d}t}\phi_{12}
\end{eqnarray}
where we define the renormalized capacitance and resistivity:  
\begin{eqnarray}
C_R=C + \frac{\hbar}{4e R}\frac{f e}{2I_{\rm c} m \overline{r}}
\quad \quad \quad 
\frac{1}{R_R}=\frac{1}{R}
+\frac{f e}{4 m \overline{r}}
\end{eqnarray}
In the frequency domain, we obtain the response function:
\begin{eqnarray}
 \phi_{12,\omega}=\frac{C 2m\overline{r}i\omega a_\omega /e }
 {-C_R \hbar \omega^2 /2e +\hbar i\omega /2e R_R +I_{\rm c}}
 \\
 \delta \phi_{\omega}=\frac{-C \hbar \omega^2/2e} 
 {-C_R \hbar \omega^2 /2e +\hbar i\omega /2e R_R +I_{\rm c}}\frac{f a_\omega}{I_{\rm c}}
\end{eqnarray}
In the limit of high-frequency limit, we recover the formulae given in the main text. In the low frequency limit we obtain: $\delta \phi_{\omega}=-\frac{C \hbar \omega^2 f a_\omega}{2e I^2_c} \cong -\omega^2/\omega^2_0 f a_\omega/I_{\rm c}$. We determine $I_{\rm max}\cong 2 I_{\rm c} [1- (\delta \phi)^2/16]$. For intermediate frequency, the renormalized resistivity  term is dominant and we find the approximation: $\delta \phi_{\omega} \cong 4 i\omega m \overline{r} a_\omega /\hbar \omega_0^2$. Using the values $I_c=nA$, $\omega=1MHz$, $a_\omega=10^3 m.s^2$, $\overline{r}=1cm$, we require a value of $\omega_0=\sqrt{I_c 2e/\hbar C}=1MHz$ imposing a minimum value of the capacitance of $C \sim 1 \mu F$ for an observable phase shift.

\subsubsection{Influence of the lattice}

In the case of lattice distortion, we  renormalize the acceleration to $a \rightarrow a_{\rm eff}=Z a$ with $Z$ determined from the various predictions in~\cite{Dessler1968}.  
According to~\cite{Schiff1970}, the role played by the lattice is controversial since there are various model descriptions with different assumptions whose validity is difficult to assess and further experimental investigations are needed to determine what is actually happening. Our work provide an opportunity to review these old theoretical approaches.

Importantly, note that in the BCS theory of superconductivity, the pairing of electron is caused by the induced lattice distortion which allows overcoming the repulsive Coulomb interaction among electrons, resulting in an effective attractive electron-electron interaction. So the lattice effects are somehow already taken into account in comparison to a normal metal.

\subsection{Phase shift from the Landau-Ginzburg model}
The definition of phase shift differs if we use the static Landau-Ginzburg approach, where the spatial dependence is of relevance instead of the time-dependence used in our approach. Yet, it is also possible to use the Landau-Ginzburg theory to achieve the same result. Indeed, for the case of a rectangle configuration with $b=0$ in each arm, the average potential due to acceleration is $\pm mac/2$. In each arm the phase difference is related to the momentum times the distance through:

\begin{eqnarray}
{m^* \mathbf v}_\pm = \hbar \nabla \phi_\pm - q^*{\mathbf A}_\pm=
{m^* \mathbf v} + \hbar \delta  \nabla \phi_\pm
\end{eqnarray}
where $v$ is the effective velocity (in our notation) and $\delta  \nabla\phi_\pm$ accounts for the extra phase gradient that should compensate the average potential. Since the sum of the kinetic energy excess and average potential energy is equal in each arm (since Cooper pair energy has to be identical  in each arm), we obtain:
\begin{eqnarray}
{\mathbf v}. \hbar \delta \nabla   \phi_+- ma c/2= 
{\mathbf v}. \hbar \delta \nabla   \phi_- + ma c/2
\end{eqnarray}
Assuming a constant phase gradient over the arm length $2c$ we find ${\mathbf v}. \delta \nabla   \phi_\pm = |{\mathbf v}|\delta  \phi_\pm/2c$ and  deduce the phase difference
\begin{eqnarray}
\delta \phi = 
\delta  \phi_+ - \delta \phi_-=\frac{2 mgc^2 a}{\hbar v}
\end{eqnarray}
which corresponds to Eq.~(4) in the main text. If $\phi_l$ is the phase change at each Josephson junction, the quantization condition imposes: 
$\delta \phi +\sum_{l=1}^N \phi_l= 2\pi k $ where $k$ is an integer. Using  $k=0$, we recover the result Eq.~(1) in the main text.

\end{widetext}

\end{document}